\documentclass[manuscript]{acmart}

\AtBeginDocument{%
  \providecommand\BibTeX{{%
    \normalfont B\kern-0.5em{\scshape i\kern-0.25em b}\kern-0.8em\TeX}}}

\usepackage{blindtext}
\usepackage{tabularx}
\usepackage{multirow}
\usepackage{graphicx}
\usepackage{subcaption}
\usepackage{dcolumn}
\usepackage{rotating}
\usepackage{xspace}
\begin{document}


\title{Designing AI Systems that Augment Human Performed vs. Demonstrated Critical Thinking}
\begingroup
\renewcommand\thefootnote{}\footnote{
 This paper was presented at the 2025 ACM Workshop on Human-AI Interaction for Augmented Reasoning (AIREASONING-2025-01). This is the authors’ version for arXiv.}
\endgroup

\author{Katelyn Xiaoying Mei}
\email{kmei@uw.edu}
\affiliation{%
  \institution{University of Washington}
  \streetaddress{}
  \city{Seattle}
  \state{WA}
  \country{USA}
  \postcode{}
}

\author{Nic Weber}
\email{nmweber@uw.edu}
\affiliation{%
  \institution{University of Washington}
  \streetaddress{}
  \city{Seattle}
  \state{WA}
  \country{USA}
  \postcode{}
}

\renewcommand{\shortauthors}{Mei et al.}



\received[accepted]{24 March 2025}

\begin{abstract} 
The recent rapid advancement of LLM-based AI systems has accelerated our search and production of information.  While the advantages brought by these systems seemingly improve the performance or efficiency of human activities, they do not necessarily enhance human capabilities. Recent research has started to examine the impact of generative AI on individuals' cognitive abilities, especially critical thinking. Based on definitions of critical thinking across psychology and education, this position paper proposes the distinction between \textit{demonstrated} and \textit{performed} critical thinking in the era of generative AI and discusses the implication of this distinction in research and development of AI systems that aim to augment human critical thinking.
\end{abstract}

\maketitle
\section{Introduction}
Generative AI (GenAI) powered by large language models (LLMs) has boosted productivity in tasks such as writing, coding, information analysis, and decision-making. However, these purported gains in efficiency do not always result in better task outcomes or human performance. In addition, the use of these systems could potentially undermine human capabilities as people develop an overreliance on AI systems for tasks ranging from simple to complex~\cite{klingbeil2024trust,cui2024assessing}. For example, while tools like GitHub Copilot may increase coding quality and efficiency~\cite{weber2024significant}, the mechanism of autocomplete does not inherently enhance individuals' programming skills or understanding~\cite{prather2023s}. Similarly, while ChatGPT and similar LLMs can increase the speed and improve the fluency of writing deliverables, they do not necessarily improve individuals' inherent writing abilities and potentially resulting in biased outcomes and content lack of depth\cite{STADLER2024108386,jakesch2023co}. 

Amidst these findings, we consider the current landscape of AI tools reveals a concerning pattern: the design of AI systems is often output-driven yet overlooks its impact on human cognitive capabilities. Recent research has started to pay attention to how generative AI (GenAI) impacts individuals' critical thinking and researchers have found mixed findings~\cite{lee2025impact,chang2025generative,gonsalves_generative_2024,lopez2024,liu2024effects}. Based on this line of work, this position paper proposes a distinction between \textit{performed} critical thinking and \textit{demonstrated} critical thinking. 
In the context of human-AI interaction, we define \textit{performed critical thinking} as cognitive process undertaken by humans independent of AI assistance. In contrast, \textit{demonstrated critical thinking} refers to either the process or product of critical thinking that occurs through the assistance of or collaboration with generative AI systems. 

These distinctions are important since the choice of interpretation could lead to different implications for i) research that examines the impact of generative AI on critical thinking and ii) divergent design objectives for AI systems to augment human critical thinking. We identify that current evaluation on the impact of GenAI on critical thinking skills often focuses on \textit{demonstrated} rather than \textit{performed} critical thinking and provide considerations for future research designs. Drawing insight from the \textit{intellect augmentation} framework conceptualized by Douglas Engelbart~\cite{engelbart2023augmenting}, we argue that while current AI systems and interaction paradigms may enhance the \textit{demonstrated} critical thinking, they do not necessarily improve the \textit{performed} critical thinking. We also discuss design considerations to accomplish the latter.

\section{Definition and Evaluation of Critical Thinking}
Critical thinking has been conceptualized through various frameworks over decades of research in education and psychology~\cite{ennis1962concept,scriven1987critical,bloom1956handbook}. Its definition varies from a general definition to a procedural process. \citet{ennis1962concept} describes critical thinking as the \textit{correct assessing of statements}.
\citet{scriven1987critical} define it as a ``a disciplined process that actively and skillfully conceptualize, apply, analyze, synthesize, and evaluate information gathered from or generated by observation, experience, reflection, reasoning or communication, to guide one's belief and action''.
Similarly focusing on the process, \citet{bloom1956handbook} identifies six cognitive activities that occur during critical thinking: \textit{knowledge}, \textit{comprehension}, \textit{application}, \textit{analysis}, \textit{synthesis}, and \textit{evaluation}. The definitions by \citet{ennis1962concept} and \citet{scriven1987critical} align with what we term \textit{performed} critical thinking—independent cognitive processes without external assistance. Yet all components within the definition by \citet{bloom1956handbook} can be either performed independently or demonstrated through human-AI collaboration. 

Researchers in psychology and education have adopted various quantitative and qualitative approaches to evaluate critical thinking~\cite{Elson2018CriticalAT,cui2024assessing}. Via a quantitative approach, researchers develop standardized tests or instruments~\cite{facione2001california,gerdts2022educational,follman1996inventory,watson1980watson}. In qualitative assessment, researchers such as open-ended questions and augmentative writing, as well as online discussion~\cite{white2011novel,stapleton2001assessing}. Across these tools, some tend to measure critical disposition~\cite{facione2001california}---the tendency to think critically---whereas the others measure performance or application of critical thinking~\cite{watson1980watson}. To examine the latter, \citet{watson1980watson} asks students to read scenarios of statements and arguments and make judgments on specific inferences, assumptions, deductions, conclusions based on each scenario. Since these traditional assessment are designed to evaluate individuals' capabilities in situations where AI assistance is not present, we consider that they measure \textit{performed} critical thinking. However, qualitative assessment could be used to measure \textit{demonstrated} critical thinking if we evaluate individuals' responses to open-ended questions and argumentative writing that are produced together with AI assistance. 

  

\begin{table}[ht]
    \centering
    \renewcommand{\arraystretch}{1.3}
    \caption{Cognitive activities defined in Bloom’s taxonomy \cite{lee2025impact,bloom1956handbook}.}
    \begin{tabular}{lp{10cm}}
        \toprule
        \textbf{Cognitive activity} & \textbf{Description} \\
        \midrule
        Knowledge & Recognition or recall of ideas, material, or phenomena \\
        Comprehension & Organising, summarising, translating, generalising, giving descriptions, and stating the main ideas \\
        Application & Using acquired knowledge to solve problems in new situations \\
        Analysis & Examining and breaking information into component parts, determining how the parts relate to one another, identifying motives or causes, making inferences, and finding evidence to support generalisations \\
        Synthesis & Building a structure or pattern from diverse elements; putting parts together to form a whole or bringing pieces of information together to form a new meaning \\
        Evaluation & Presenting and defending opinions by making judgements about information, the validity of ideas, or quality of work based on a set of criteria \\
        \bottomrule
    \end{tabular}
\end{table}

\section{Evaluating the Impact of Generative AI on \textit{Performed} vs. \textit{Demonstrated} Critical Thinking}
Recent research raises questions about how generative AI affects individuals' critical thinking abilities~\cite{lee2025impact,chang2025generative,gonsalves_generative_2024,gerlich_ai_2025}. Current research often conflates \textit{performed} and \textit{demonstrated} critical thinking by primarily assessing the latter rather than the former. For example, recent studies that suggest the use of GenAI improves critical thinking evaluate only the final outputs (e.g., written responses) that participants produce with AI assistance~\cite{chang2025generative,gonsalves_generative_2024}. Such methodologies fail to measure whether individuals' \textit{performed} critical thinking has improved. We encourage future research to adopt methodologies that can distinctly evaluate both forms of critical thinking. To robustly assess the impact of AI on critical thinking abilities, researchers should implement study designs that assess participants' \textit{performed} critical thinking. For example, several studies have used quasi-experiments or interventions to evaluate whether the usage of ChatGPT impacts students' critical thinking abilities where students' critical thinking is assessed before and after the interventions~\citet{lopez2024,liu2024effects}. \citet{cui2024assessing} conduct a qualitative assessment of students' dialogue in a classroom setting. Future research should also consider longitudinal frameworks that track changes in independent cognitive abilities over time. 

\section{Augmenting \textit{Performed} vs. \textit{Demonstrated} Critical Thinking}

Designing technology to augment human capabilities has been one of the major objectives of technology development. In 1962, \citet{engelbart2023augmenting} presents the intellect augmentation framework to delineate how we can design artifacts to augment human capabilities. He considers augmentation is achieved if humans are more efficient in finding better solutions to complex problems. \citet{engelbart2023augmenting} indicates that there are \textit{processes} (''little steps or actions'') that take place in humans' problem solving.
 
Intellect augmentation can be accomplished by making any of these processes efficient and better.
It is worth noting that within his framework, intellect augmentation is not of human but of a system where humans with \textit{augmentation means}---which he referred as ``\textit{H-LAM/T}''\textit{system} (Humans using Language, Artifacts, and Methodology). Detailed description of each component is included in \autoref{tab:augmentation_means}. 

While Engelbart's framework provides an insightful road map for intellect augmentation, the rapid evolution of AI and intelligent technologies has necessitated expanding these foundational ideas. \citet{xia_design_2013} provides a new interpretation of intellect that goes beyond problem solving, extending it to other cognitive domains of human capabilities such as memory, motivation, and decision making. Building upon the intellect augmentation framework, they derive steps for designing artifacts that support augmentation of different cognitive domains: 1) \textit{consider the desired state after augmentation}; 2) \textit{identify the processes for the task}; 3) \textit{identify how artifacts can change a process or processes}. 

Based on these augmentation frameworks, we interpret what it means to augment human critical thinking via AI systems. To augment critical thinking, the distinction between \textit{performed} and \textit{demonstrated} critical thinking presents different design paths for AI systems. \textbf{Augmenting demonstrated critical thinking} focuses on the quality of the final output or collaborative process that is associated with individuals' critical thinking. \textbf{Augmenting performed critical thinking} emphasizes the improvement of independent critical thinking activity after the interaction with AI systems. Building upon the definition of \citet{bloom1956handbook} for critical thinking, augmentation of \textit{demonstrated} critical thinking can be accomplished by improving \textbf{the final output} associated with any of the components (e.g. \textit{knowledge, comprehension, application, analysis, synthesis, evaluation}). For example, current conversational AI systems can explain complex concepts to individuals, enabling more efficient and rapid comprehension, thus augmentation is accomplished. Yet it does not necessarily augment individuals' \textit{performed} comprehension skill. On the contrary, overrelying on AI assistance to comprehend complex concepts may negatively impact individuals' independent comprehension capability as they practice it less on their own.

Both forms of augmentation serve important but distinct purposes in human-AI collaboration. Augmenting demonstrated critical thinking may be particularly valuable in time-sensitive contexts, collaborative workflows, or when addressing highly complex problems that exceed individual cognitive capacity. Meanwhile, augmenting performed critical thinking becomes essential for long-term skill development, educational settings, and maintaining human autonomy in reasoning. To cultivate  \textit{performed} critical thinking, AI systems need to train and empower users to practice high-quality critical thinking independently. We encourage future research to draw insight from research in education and learning to identify effective interactive paradigms or features that scaffold critical thinking processes rather than simply delivering conclusions. For example, numerous work in education research identify the effective use of questioning, classroom debates, and writing could promote critical thinking in students \cite{walker2003active}. Future research could explore similar strategies---such as providing structured frameworks, guiding questions without resolving the analytical process completely--- could support users in developing their own critical thinking skills.


\begin{table}[h]
    \centering
       \caption{Augmentation Means within the Intellect Augmentation Framework by \citet{engelbart2023augmenting}}
    \renewcommand{\arraystretch}{1.2}
    \begin{tabular}{ l|p{10cm}}
        \hline
        \textbf{Augmentation Means} & \textbf{Description} \\ 
        \hline
        Artifacts & Physical objects designed to provide for human comfort, for the manipulation of things or materials, and for the manipulation of symbols. \\ 
        \hline
        Language & The way in which the individual parcels out the picture of his world into the concepts that his mind uses to model that world, and the symbols that he attaches to those concepts and uses in consciously manipulating the concepts ("thinking"). \\ 
        \hline
        Methodology & The methods, procedures, strategies, etc., with which an individual organizes his goal-centered (problem-solving) activity. \\ 
        \hline
        Training & The conditioning needed by the human being to bring his skills in using Means 1, 2, and 3 to the point where they are operationally effective. \\ 
        \hline
    \end{tabular}
 
    \label{tab:augmentation_means}
\end{table}

\section{Conclusion}
The rapid development and integration of AI systems in our everyday lives raises questions not only for their impact on human cognitive abilities but also for established concepts and methodologies in research. Focusing on critical thinking, this position paper identifies the need to further distinguish types of critical thinking in the era of AI: \textit{performed} versus \textit{demonstrated} critical thinking. Via such distinction, we discuss the nuances in current research findings regarding the impact of AI systems on human critical thinking, calling for explicit considerations of this distinction in future studies. More importantly, such a distinction implicates our design of AI systems to augment critical thinking. We encourage researchers and developers to keep this distinction in mind as systems that augment \textit{demonstrated} critical thinking are not the same as those that augment \textit{performed} critical thinking.

\bibliographystyle{ACM-Reference-Format}
\bibliography{reference}

\end{document}